\documentclass[letterpaper, 10 pt, conference]{ieeeconf}  
\IEEEoverridecommandlockouts                              
\usepackage{graphicx} 
\usepackage{array} 
\usepackage{multirow} 
\usepackage{amsmath}
\usepackage{graphics} 
\usepackage{epsfig} 
\usepackage{mathptmx} 
\usepackage{times} 
\usepackage{amssymb}  
\usepackage{xcolor} 
\usepackage{tabularx} 
\usepackage{colortbl}
\DeclareMathOperator*{\argmax}{arg\,max}

\DeclareMathOperator*{\GaussianDist}{Gaussian\,Dist}
\DeclareMathOperator*{\SamplingDist}{Sampling\,Dist}

\title{\LARGE \bf
Anatomical Region Recognition and Real-time Bone Tracking Methods by Dynamically Decoding A-Mode Ultrasound Signals}

\author{Bangyu Lan$^{1}$, Stefano Stramigioli$^{1}$ and Kenan Niu$^{1}$ 
\thanks{$^{1}$Robotics and Mechatronics, University of Twente, Enschede, AE, The Netherlands}%
}

\begin{document}

\maketitle
\thispagestyle{empty}
\pagestyle{empty}

\begin{abstract}
Accurate bone tracking is crucial for kinematic analysis in orthopedic surgery and prosthetic robotics. Traditional methods (e.g., skin markers) are subject to soft tissue artifacts, and the bone pins used in surgery introduce the risk of additional trauma and infection. For electromyography (EMG), its inability to directly measure joint angles requires complex algorithms for kinematic estimation. To address these issues, A-mode ultrasound-based tracking has been proposed as a non-invasive and safe alternative. However, this approach suffers from limited accuracy in peak detection when processing received ultrasound signals. To build a precise and real-time bone tracking approach, in this paper, a deep learning-based method was introduced for anatomical region recognition and bone tracking using A-mode ultrasound signals, specifically focused on the knee joint. The algorithm is capable of simultaneously performing bone tracking and identifying the anatomical regions where the A-mode ultrasound transducer was placed. It contains the fully connection between all encoding and decoding layers of the cascaded U-Nets to focus only on the signal region that is most likely to have the bone peak, thus pinpointing the exact location of the peak and classifying the anatomical region of the signal. The experiment showed a 97\% accuracy in the classification of anatomical regions and a precision of around 0.5$\pm$1mm for tracking movements of the various anatomical areas surrounding the knee joint. In general, this approach shows great potential beyond the traditional method, in terms of the accuracy achieved and the recognition of the anatomical region where the ultrasound has been attached as an additional functionality.
\end{abstract}

\section{INTRODUCTION} 
Bone tracking technology is essential for the kinematic analysis of human body, particularly in the knee joint. The highly precise tracking produces accurate kinematics data, vital for surgical procedures \cite{li2022accuracies}, prosthetic robotics \cite{embry2020analysis}, and wearable exoskeletons \cite{meier2023evaluating}. Typically, the gold standard of tracking is achieved by using bone pins with optical markers \cite{niu2018feasibility}, but it introduces invasive procedures and infection risks to subjects. Another method is electromyography (EMG)-based techniques \cite{lotti2020adaptive, 10304809, 6377275, 10304819}, but indirect measurement based on muscle activation patterns requires complex algorithms to analyze kinematics. In this context, a more accurate and convenient approach is preferable to obtain the knee kinematics in a non-invasive manner. 

\begin{figure} [t!]
    \centering
    \includegraphics[width=1\linewidth]{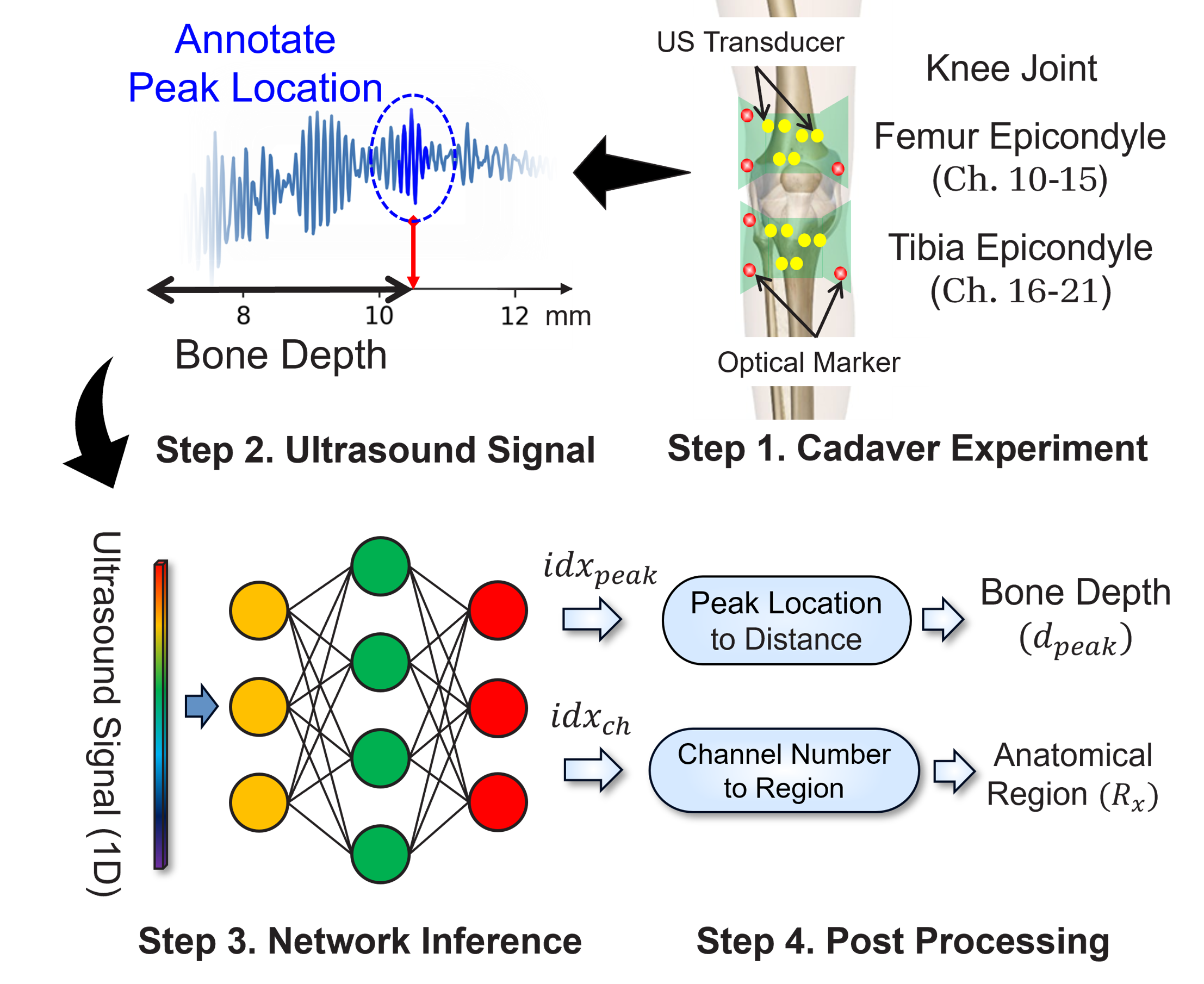}
    \caption{Steps to build our method: A cadaver experiment to get the dataset and annotate the bone peak location. Our network is trained by the dataset and infers the actual bone depth and anatomical region.}
    \label{fig1}
\end{figure}

Recently, an A-mode ultrasound (US) based tracking method has been introduced as a solution \cite{niu2018novel}. Compared to B-mode US, A-mode US can perform bone tracking in real-time, without the receiving and processing time of 2D images, and the need to analyze medical images by experts. Compared with other tracking techniques (e.g., bone pins, skin markers \cite{terlemez2014master}, fluoroscopy \cite{8756013}, MRI \cite{zeineldin2021hybrid}, etc.), A-mode ultrasound combines advantages of safety, high accuracy, non-invasiveness, and cost-effectiveness. However, its accuracy and robustness are compromised due to the reliance on traditional peak detection to analyze one-dimensional raw US signals. Traditional A-mode US methods (\cite{niu2018feasibility,niu2018measuring,niu2018novel,niu2018situ}) used signal processing theory to detect the highest peak (in a pre-defined local range) representing the precise bone distance from skin, which have not considered actual bone peak profiles, and heavily rely on the expert knowledge of approximate local range. This paper tried to solve the issues using a deep learning framework.

In related research fields, deep learning has been employed for signal peak recognition. For example, in the diagnosis of heart disease, a U-Net framework was developed to segment and identify meaningful peaks in raw EEG signals \cite{moskalenko2020deep}. However, the A-mode US signal presents a unique challenge: the meaningful bone peaks are actually sparse and ambiguous due to acoustic strength attenuation or the unclear interface between the soft tissues (e.g. tendon and muscles) and the bony surface. For this reason, to our knowledge, few studies have reported using deep learning for A-mode ultrasound diagnosis in knee kinematics tracking or real-time bone tracking. A novel method that considers both the local features (sparse bone peak) and the general features (entire ultrasound echo signal) of A-mode US signals could be a potential solution to address this challenge.

In this study, we focused on tracking the knee joint, which poses significant challenges for traditional tracking methods due to the complex curvature of the skin and bone surfaces around the femur epicondyle and the tibia epicondyle. These anatomical features create difficulties in achieving accurate distance measurements. The powerful capability of deep learning to extract abstract features can probably enhance the precision of peak recognition in this area. This is because bone peaks, despite their subtle and complex characteristics in geometry, may share underlying similar features that deep learning algorithms can identify and analyze. 


Specifically, we employed fully connected cascaded U-Nets with the interconnections determined by the Sampling-based Proposal (SBP), enhancing efficiency and accuracy. The SBP adopted a probabilistic way to dynamically select a much narrower signal region as the input of the refined U-Net's segmentation. Furthermore, an anatomical region classification network in the first U-Net bottleneck layer facilitated the extraction of comprehensive signal knowledge, enabling anatomical region recognition. In general, this approach not only achieved high precision in tracking knee joint movement but also extracted anatomical knowledge from the US signal simultaneously. The recognized anatomical region is especially useful after tracking the location of the bone. To have a complete bone registration, accurate predefined landmarks on the bones are required for the correct alignment between 3D scans of bony surface and US transducer locations\cite{niu2018feasibility, 6290782}. The anatomical region detected by our method can provide position calibration for the tracking robot.  

In summary, our method simultaneously identifies anatomical regions and performs real-time tracking of the bony surface. It shows great potential for usage in prosthetic robot control and bone or tissue tracking.


\begin{figure*} [t!]
    \centering
    \includegraphics[width=0.9\linewidth]{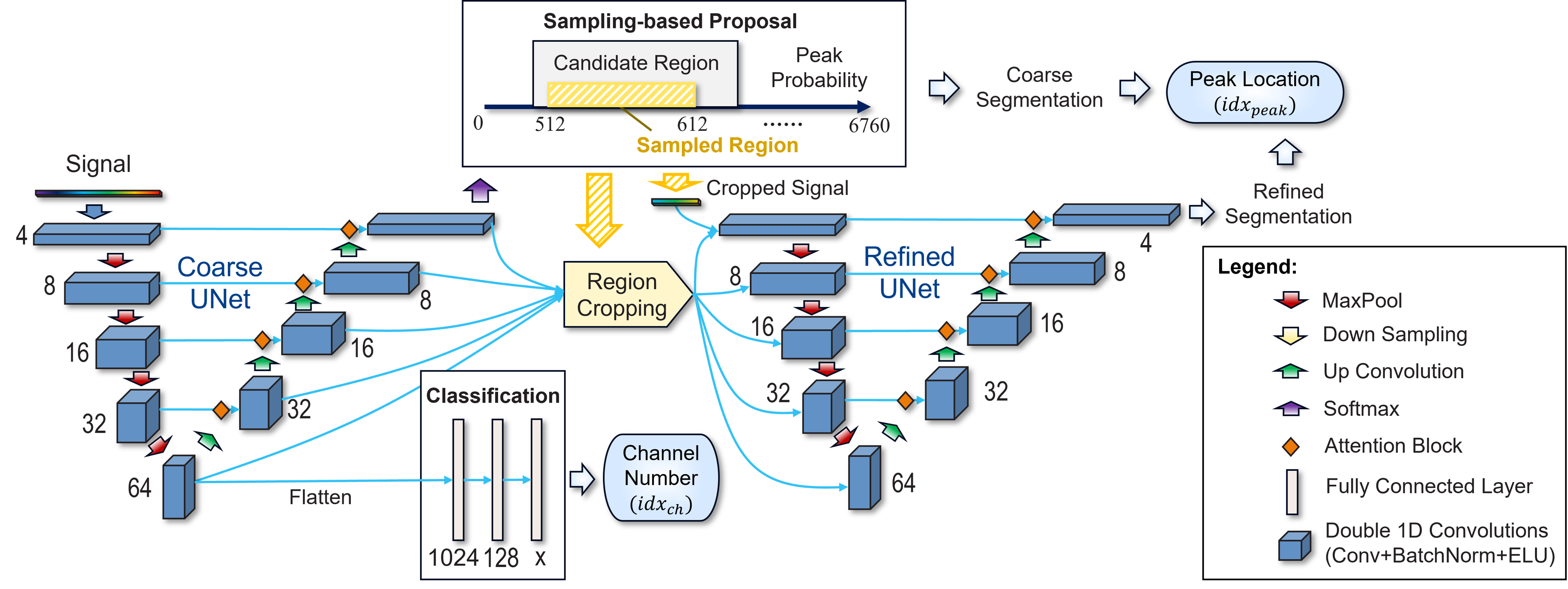}
    \caption{The proposed network had two U-Nets with different scales of perception fields (for bone peak detection) and input signal classification. The input was a 1D ultrasound signal. The outputs were the signal classification results and the prediction of the peak location. The peak location prediction was a segment after thresholding the predicted probability sequence. The two U-Nets were connected by Sampling-based Proposal.}
    \label{fig3}
\end{figure*}

\section{Method} 

\subsection{Motion Tracking System} 
Our motion tracking method began with collecting the knee joint motion dataset from a cadaver specimen in a previous work \cite{niu2018situ}. This dataset included the positions of optical markers from bone pins and US holders. Based on the dataset, in this paper, the optical markers were used to transform the actual 3D positions of US transducers (frame $\Psi_1$) and bone pins (frame $\Psi_2$) into the experimental coordinate frame $\Psi_3$. After rendering their 3D positions relative to the bony surface in the same coordinate frame, the intersections between the bony surface and the directions of the US waveform could be found. The depth of bone under the skin, denoted $d$, was obtained by calculating the distance between the intersections and the origin of the ultrasound transducers \cite{niu2018measuring}. For each transducer, the corresponding depth distance corresponded to the location of a specific bone peak in the A-mode US signal. This conversion from depth distance to the index of the bone peak ($idx_{peak}$) was described in Equation (\ref{peak_dist_conversion}), where $v=1540 \text{ m/s}$ (the speed of ultrasound in soft tissues) and $f_s=40\times10^6 \text{ Hz}$ (the ultrasound sampling rate). Here, $d_{unit}$ represents the actual unit length (in millimeters) for each unit of the total 6760-unit length (approximate 130mm) of ultrasound signals. 

\begin{equation}
\label{peak_dist_conversion}
idx_{peak}=\frac{d}{d_{\text{unit}}}=\frac{d}{\left(2 \times \frac{v}{f_s} \times 1000\right)}
\end{equation}

In the end, the signals from different anatomical positions of knee joint were collected, with each anatomical region corresponded to one ultrasound channel, where the ultrasound transducer attached. All bone peak locations in the signals from all channels of femur and tibia epicondyle were annotated. These annotations served as dataset labels to train the proposed cascaded U-Nets separately (femur epicondyle or tibia epicondyle). During testing, the network detected bone peaks in the signal, which was later converted to the actual depth of bone for evaluation using the calculated ground truth distance.


\subsection{Overall Structure of the Network} 

Our method was designed to recognize the anatomical region and perform bone tracking with high accuracy. To this end, a novel structure of fully connected cascaded U-Nets was proposed, which was depicted in Fig. \ref{fig3}. The input of Coarse U-Net was a 1D signal. The output features of the Coarse U-Net bottleneck layer were used to classify the signal channel (anatomical position). The SBP pinpointed the dynamic region that was most likely to contain bone peaks in a probabilistic way (explained below in details). The output features from all decoder layers of the Coarse U-Net were linked to the encoder of the Refined U-Net through SBP. The Refined U-Net output yielded a more precise peak detection (existing as a segment form). The combination of two segments ultimately determined the existence and location of the bone peak.
 
\subsection{Details of the Structure}
\subsubsection{Cascaded U-Nets}
Inspired by \cite{liu2019cu}, our cascaded U-Nets structure was designed to utilize the underlying hierarchical structure within the US signal. The two U-Nets shared a similar structure: each comprised an encoder and a decoder with five layers of dual 1D convolutions. The number of kernels for each convolution was indicated next to the dark blue cubes in Fig. \ref{fig3}. In the first four layers, the encoder's output was filtered by the feature from the deeper layer through the 1D feature attention block \cite{chen2022multiple} before proceeding to the decoder. This filtering suppressed irrelevant features (peaks resulting from other tissues or noise) and highlighted the salient bone peak-related features. For the Refined U-Net, the encoder's input at each layer was a concatenation between the outputs of Region Cropping and the MaxPooling. The advantage of cascaded U-Nets instead of only one U-Net was the augmented signal scale, which helps to improve the method's perceptual resolution and improve the detection accuracy.

\subsubsection{Sampling-based Proposal}
To determine the region that is most likely to have the bone peak, a Sampling-based Proposal (SBP) inspired by \cite{lan2024deep} was established between the layers of the coarse U-Net encoder and the refined U-Net encoder. The steps were as follows: Initially, the Coarse U-Net output was converted into the probability of bone peak ($p^{peak}_i$ at the $i^{th}$ location) using SoftMax, serving as a preliminary guess of the location of bone peaks. Subsequently, a candidate region (a sequence of indexes $\{idx_{start}, idx_{start+1}, ..., idx_{end}\}$) was identified around the point of highest probability, the size being three times the width of the final sampled region. Within this candidate area, a Gaussian distribution ($\GaussianDist(mean, std)$) was generated for each probability point. The cumulative effect of all these Gaussian distributions formed the final sampling distribution$\SamplingDist$, which is found in Equation (\ref{build_dist}). 

\begin{equation}
\label{build_dist}
\SamplingDist = \sum^{end}_{i=start} p^{peak}_{i}*\GaussianDist(idx_{i},1)
\end{equation}

The final signal region was then sampled using this distribution as the input of the Refined U-Net. Compared to \cite{lan2024deep}, the region of the segment sampled in SBP was also used to crop the features of each output layer of the Coarse U-Net decoder. Note that before cropping, the region was down-sampled first to match the feature resolution in the corresponding layer. The outputs of Region Cropping were directly concatenated with the inputs of each layer in the Refined U-Net. Overall, this strategy captured only the essential regions and increased the resolution. Compared to the normal network, this probabilistic approach offered a better recognition of the dynamic peak region, facilitating further investigation by the following network.

\subsubsection{Classification Network}
The classification network had three fully connected layers to reduce the dimension. The number of neurons in each layer was specified inside the Classification box (Fig. \ref{fig3}), where 'x' denoted the number of categories (3 for the femur and 5 for the tibia). LeakyReLU, with a negative slope of 0.1, was used as the activation function between layers. The final classification result was determined by the last layer with a Softmax function. The advantage of linkage between the classification network and the bottleneck layer was that the Coarse U-Net encoder had the capacity to capture the entire signal. This meant that the encoded features from the bottleneck contain comprehensive information, not just bone peaks related but also various soft tissue characteristics, which are crucial for anatomical area identification.

\subsubsection{Network Output}
The network generated two outputs, which were represented with the light blue background oval shapes in Fig. \ref{fig3}. The upper right light blue part ("Peak Location") pinpointed the precise bone peak location, using both coarse and refined segmentation results from two U-Nets. This segmentation was obtained by applying a threshold to the probability of bone peak. The rule for peak determination was based on the priorities of two segmentations: the Coarse segmentation confirmed the existence of a peak, while the Refined segmentation ascertained its precise location. Therefore, a bone peak was considered to exist only if it was indicated by the Coarse U-Net output. Once a bone peak was confirmed to exist, the Refined segmentation was used to determine the exact position. Regarding the bottom-middle light blue box, it gave the signal classification by Argmax on the output of the classification network. This was illustrated in Equation (\ref{signal_cls}), where $n$ is the total number of channels (anatomical regions), $p^{ch}_i$ is the probability of the $i^{th}$ channel.

\begin{equation}
\label{signal_cls}
R_x \leftarrow idx_{ch}= \argmax(p^{ch}_1,p^{ch}_2,...,p^{ch}_n)
\end{equation}

As mentioned before, each channel corresponded to one anatomical region. The anatomical region was characterized by unique subcutaneous tissues, creating distinctive signal characteristics that are useful for classification.

\subsection{Training Strategy and Post-Processing}
To train the network for accurate segmentation, dice loss and cross-entropy loss were used for both the Coarse U-Net $l_{dice}$, $l_{ce}$ and Refined U-Net $l'_{dice}$, $l'_{ce}$. Dice loss \cite{milletari2016v} can mitigate the problem of sparse foregrounds. This was crucial as over 6760-unit signal length, the peak region spanned merely 10 units, a dimension easily overlooked when relying solely on cross-entropy loss. Equation (\ref{dice_loss}) detailed the dice loss formula. Cross-entropy loss was used as a binary classification for the network to identify the foreground (peak region) or background at each unit. For classification, the training loss $l_{cls}$ was also the cross-entropy loss. The final training loss was in Equation (\ref{total_loss}). The network was trained by RMSprop optimization \cite{elshamy2023improving} with a learning rate of 1e-5, a batch size of 10 and a duration of 50 epoches \cite{moskalenko2020deep}.

\begin{equation}
\label{dice_loss}
Dice Loss (DL) = 1-\frac{2 \times \sum_{i=0}^n(p_i^{pred}*p_i^{true})+\epsilon}{\sum_{i=0}^np_i^{pred}+\sum_{i=0}^np_i^{true}+\epsilon}
\end{equation}

\begin{equation}
\label{total_loss}
loss = l_{dice} + l_{ce} + l'_{dice} + l'_{ce} + l_{cls}
\end{equation}

To construct the training dataset, two distinct movements of the knee joint were collected. They were merged and segmented into 2033 samples for all transducer channels. Within the femur epicondyle channels No. 10 to No. 15, only channels No. 11, No. 12, and No. 15 exhibited discernible bone peaks; the others were excluded. The signals from the viable channels were truncated if the strengths exceeded 5000. These truncated signals were augmented tenfold by shifting the units on the x-axis. Subsequently, the dataset was divided into training and testing parts in an 8:2 ratio. An identical process was also applied to the tibia epicondyle channels. We shuffled US signals from all channels in the same epicondyle for training and testing, as we assumed that the bone peak in the same area (femur or tibia) exhibited similar profiles.

During post-processing, Equation (\ref{peak_to_dist}) was used to convert the peak location to the actual depth of the bone. We also verified the anatomical region of the classified channel.

\begin{equation}
\label{peak_to_dist}
d=d_{peak}=idx_{peak} \times d_{\text{unit}}
\end{equation}

\subsection{Evaluation}
To demonstrate the improvement in accuracy, we introduced the traditional method in \cite{niu2018situ} for comparative analysis. The conventional method of detecting bone peaks involves using expert knowledge to pinpoint the general vicinity of the peak. Within this localized area, a traditional peak detection was used \cite{niu2018feasibility,niu2018measuring,niu2018novel,niu2018situ} to identify the highest peak as the bone peak.

To evaluate our approach, we first collected the bias distance between the predicted peaks (the peak position was regarded as the middle position of the segmentation) and the ground truth peaks. Then the mean and standard deviation of bias were calculated. Outliers that were much divergent from most biases were analyzed by examining the corresponding 3D position of the knee joint and the US waveform, which was shown in Fig. \ref{fig5}. We also recorded the network inference time to determine the speed of our method.

\begin{table}[t!]
\caption{peak detection accuracy and proportion of the sub-millimeter bias. Epi. refers to epicondyle}
\centering
\begin{tabular}{|>{\centering\arraybackslash}m{0.6cm}|>{\centering\arraybackslash}m{1.0cm}|>{\columncolor{gray!20}}c|>{\centering\arraybackslash}m{1.7cm}|>{\centering\arraybackslash}m{0.9cm}|}
\hline
    \textbf{Area} & \textbf{\begin{tabular}{@{}c@{}}\textbf{Channel} \\ \textbf{(Region)}\end{tabular}} & \textbf{Traditional (mm)} & \textbf{Cascaded UNets (mm)} & \textbf{\% sub-mm}\\
\hline
\multirow{3}{*}{\textbf{\begin{tabular}{@{}c@{}}\textbf{Femur} \\ \textbf{Epi.}\end{tabular}}} & 11 $(R_\alpha)$ & 1.455 $\pm$ 1.494 & 0.434 $\pm$ 0.843 & 84.7\% \\
\cline{2-5}
& 12 $(R_\beta)$ & 2.334 $\pm$ 2.141 & 0.453 $\pm$ 1.201 & 89.7\% \\
\cline{2-5}
& 15 $(R_\gamma)$ & 3.276 $\pm$ 3.183 & 0.551 $\pm$ 1.322 & 84.2\% \\
\hline
\multirow{5}{*}{\textbf{\begin{tabular}{@{}c@{}}\textbf{Tibia} \\ \textbf{Epi.}\end{tabular}}} & 16 $(R_\delta)$ & 2.778 $\pm$ 2.111 & 0.582 $\pm$ 1.205 & 87.8\% \\
\cline{2-5}
& 17 $(R_\epsilon)$ & 2.808 $\pm$ 1.356 & 0.604 $\pm$ 0.750 & 87.1\% \\
\cline{2-5}
& 18 $(R_\zeta)$ & 2.033 $\pm$ 1.472 & 0.666 $\pm$ 0.852 & 77.7\% \\
\cline{2-5}
& 19 $(R_\eta)$ & 4.686 $\pm$ 1.477 & 0.312 $\pm$ 0.662 & 89.7\% \\
\cline{2-5}
& 20 $(R_\theta)$ & 3.314 $\pm$ 1.907 & 0.683 $\pm$ 0.959 & 77.9\% \\
\hline
\end{tabular}
\label{table1}
\end{table}


\section{Result} 
TABLE \ref{table1} is the quantitative results of the bias distance. The dark background cells referred to the traditional method's results. It showed that our method achieved an approximately $0.5 \pm 1$mm the accuracy. In contrast, the traditional method using \cite{niu2018situ} showed the average accuracy of only $2.835 \pm 1.893$mm. In addition, the cascaded UNets achieved high classification accuracies, with 97.04\% for classifying three channels signals in femur epicondyle, and 97.48\% for classifying five channels signals in tibia epicondyle.

\begin{figure*} [t!]
    \centering
    \includegraphics[width=0.95\linewidth]{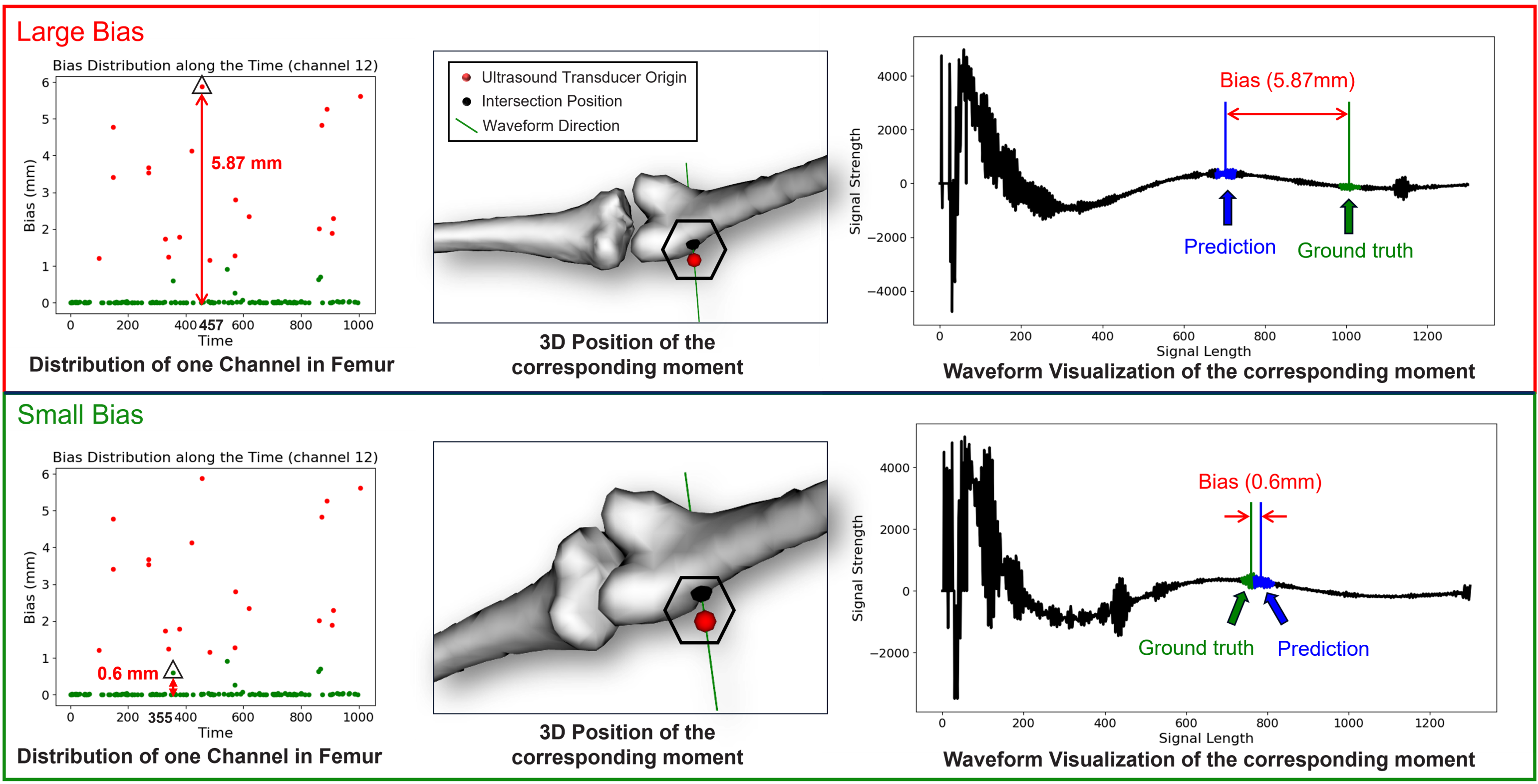}
    \caption{Closer look at the large and small bias. The distribution of bias along the time (across the entire 1017 samples) is plotted on the left. The 3D position of knee joint in the middle was the same moment of the specified bias. In the right waveform figure, the bias was visually showed.}
    \label{fig5}
\end{figure*}

For an in-depth examination of the outliers, Fig. \ref{fig5} presented two situations that contained both a large and a small bias situations in channel 12. In the top row, an outlier was carefully analyzed in which the prediction deviated by 5.87mm from the ground truth location. The corresponding 3D position and waveform at this moment were also presented. Similarly, plots representing low error scenarios are provided in the bottom row, offering a balanced view to investigate the method's performance.

Except for these experiments, the network inference speed is recorded, which is 15 ms per batch using the normal laptop (Intel i7-10875H CPU, GeForce RTX 2080 Super Max-Q designed GPU, and 32GB RAM, 2TB SATA SSD). This means that our method has a rapid peak detection speed.

\section{Discussion}
In this paper, we proposed a deep learning based method that utilized A-mode ultrasound (US) for measuring bone depth and identifying anatomical regions during bone tracking. The network could process data in 15ms per signal, which meant that our method is capable of processing US signal in real-time while obtaining anatomical regions as extra information.

Our method has been improved upon the CasAtt-UNet \cite{lan2024deep} by fully integrating cascaded dual U-Nets in each layer and modifying Sampling-based Proposal (SBP) structure for an end-to-end training. This enhancement streamlined the learning process and simplified training. Additionally, the integration of a classification network offered deeper insights into the 1D signal and validates the discriminating features learned by the Coarse U-Net encoder. With the ability to identify anatomical regions, our method offers an option for position calibration, which is critical for precise landmark registration during post-processing in bone tracking \cite{8756013, henrich2024registered}. This feature is particularly beneficial in computer-assisted and robotic-assisted orthopedic surgeries, such as Total Knee Arthroplasty (TKA) \cite{mannan2018increased}, where accurate bone tracking is vital for kinematic analysis and disease diagnosis. Furthermore, in applications like exoskeletons \cite{rinaldi2023flexos} and Human-Robot Interaction \cite{10123352}, accurate position calibration is essential to ensure correct location sensing, thus facilitating task completion.

In addition, our study had a limitation due to the use of a single cadaver specimen. Inclusion of specimens with varied human characteristics such as gender and age would enhance the robustness of our results. To mitigate this limitation, we gathered two datasets that were recorded at different postures and times of the day. In data augmentation, we pre-processed these datasets by slicing and shuffling them to demonstrate the validity and generalizability of our approach.

When looking at TABLE \ref{table1}, the bias between the predicted and ground truth peaks was around $0.5 \pm 1$mm, indicating that the proposed method achieved a sub-millimeter accuracy for most cases, significantly surpassing the accuracy of the traditional method in the dark cells, derived from \cite{niu2018situ}. In addition, there are slight variations across channels, possibly due to the different characteristics of the soft tissue surrounding the knee joint. However, these variations also provided unique signal characteristics beneficial for classification, resulting in a high accuracy rate of 97\%. This highlights the efficacy of our anatomical-aware bone tracking approach, finely classifying anatomical regions during bone peak detection.

To have a closer look at the large and small errors that occurred in the femur epicondyle, we conducted a detailed analysis of the scenarios in channel 12, with the findings presented in Fig. \ref{fig5}. In the 3D position of the top row (Large Bias), we observed that at the $457^{th}$ moment, the right leg was transitioning from extension to flexion. During this phase, the labeled ground truth was situated in the green segment of the waveform. Probably the peak profile in the predicted location resembling the actual bone peaks in the training dataset, the network mistakenly identifies the bone peak. Notice that there were several possible reasons for the attenuated bone peaks: (1) the curvature of the skin at that moment could lead to loss of skin contact with the transducers, leading to incorrect ground truth calculation and labeling, and (2) the specific posture of the specimen (fixed on the surgical table) might cause an unusual distribution of the soft tissues, attenuating the bone peaks. However, these potential causes should be investigated in the future works. 

In contrast, the bottom row of Fig. \ref{fig5} (Small Bias) illustrates a case of small error, where the leg was in an extension position. In the vicinity of the waveform, the bone peak has an apparent shape without other similar-strength peaks presented nearby to interfere with peak recognition. This makes it easier to accurately identify the bone peak in the signal.

In the future, this technique will be transferred and adopted in an in-vivo settings. In clinical practice, the bone-pins insertion for getting the ground truth labels is impossible due to the invasiveness. An alternative plan was to use other techniques (e.g., B-mode US \cite{jonkergouw2024application}, skin markers \cite{terlemez2014master}) together to check the position of bony surfaces more precisely, and mark the approximate bone peaks in US signals to train the network. The further bone registration after using anatomical landmarks \cite{niu2018novel} can continually minimize the bone tracking errors. Additionally, since the bone peak profiles for in-vivo are different from the cadaver's, new experiments using in-vivo's data for training network are required to validate the accuracy of the approach, which is crucial to evaluate generalizability and robustness of the method in clinical practice.

\section{CONCLUSIONS}
In this study, an anatomical region perceivable method capable of real-time bone tracking was proposed. This method significantly exceeds the accuracy of traditional techniques using A-mode US for bone measurement. Additionally, it demonstrates a high precision of anatomical region identification. Our approach makes the A-mode ultrasound a safe and non-invasive alternative for tracking bone movements and identifying anatomical region. Potentially, our approach not only enhances current capabilities of A-mode ultrasound but also paves the way for its future integration into robotics and prosthetic systems, promising advancements in accurate kinematics measurements that provide real-time feedback for precise robotics control.


\bibliographystyle{IEEEtran}
\bibliography{IEEEexample}

\end{document}